\documentclass[preprint]{aastex} 

\slugcomment{MNRAS: in press} 
 
\shorttitle{Collimated outflows in the planetary nebula IC\,4846} 

\shortauthors{Miranda et al.} 
 
\begin{document} 

\title{Precessing collimated outflows in the \\ planetary 
nebula IC\,4846} 

\author{Luis F. Miranda}
\affil{Instituto de Astrof\'{\i}sica de Andaluc\'{\i}a, CSIC, Apdo. Correos 
3004, E-18080 Granada, Spain}
\email{lfm@iaa.es}

\author{Mart\'{\i}n A. Guerrero}
\affil {Department of Astronomy, University of Illinois at Urbana-Champaign 
1002 West Green Street, Urbana, IL 61801 USA}
\email{mar@astro.uiuc.edu}

\author{Jos\'e M. Torrelles}
\affil{Institut d'Estudis Espacials de Catalunya (IEEC/CSIC) and Instituto de 
Ciencias del Espacio (CSIC), Edifici Nexus, C/ Gran Capit\'a 2-4, E-08034 
Barcelona, Spain}
\email{torrelles@ieec.fcr.es}

\newcommand{\ap}{\mbox{a.p.}}                   
\newcommand{\arcmper}{\mbox{$\rlap.{' }$}}      
\newcommand{\arcsper}{\mbox{$\rlap.{'' }$}}     
\newcommand{\rlas}{\mbox{$\rlap.{'' }$}}        
\newcommand{\cc}{\mbox{cm$^{-3}$}}              
\newcommand{\tco}{\mbox{$^{13}$CO}}                             
\newcommand{\cdo}{\mbox{C$^{18}$O}}                             
\newcommand{\dec}{\mbox{$\delta(1950)$}}        
\newcommand{\dgr}{\mbox{$^\circ$}}              
\newcommand{\dper}{\mbox{$\rlap.^{\circ}$}}     
\newcommand{\et}{et al.\ }                      
\newcommand{\gapprox}{\mbox{$_>\atop{^\sim}$}}  
\received{\today}
\accepted{}

\begin{abstract}

We present [N~{\sc ii}] and H$\alpha$ images and high resolution long--slit 
spectra of the planetary nebula IC\,4846, which reveal, for the first time, 
its complex structure and the existence of collimated outflows. The object 
consists of a moderately elongated shell, two (and probably three) pairs of 
collimated bipolar outflows at different orientations, and an attached 
circular shell. One of the collimated pairs is constituted by two 
curved, extended filaments whose properties indicate a high velocity, bipolar 
precessing jet. A difference of $\simeq$ 10 km\,s$^{-1}$ is found 
between the systemic velocity of the precessing jets and the centroid 
velocity of the nebula, as recently report for Hu\,2-1. We propose that 
this difference is due to orbital motion of the ejection source in a binary 
central star. The orbital separation of $\leq$ 30 AU and period $\leq$ 100 
yr estimated for the binary are similar to those in Hu\,2-1, linking the 
central stars of both planetary nebulae to interacting binaries. 
Extraordinary similarities also exist between IC\,4846 and the bewildering 
planetary nebula NGC\,6543, suggesting a similar formation history for both 
objects.

\end{abstract}


\section{Introduction}

IC\,4846 (PN\,G\,027.6$-$09.6) is a compact planetary nebula (PN) 
whose morphology has not been studied in detail yet. The only available 
information on its structure is provided by the VLA 6\,cm continuum 
observations by Kwok (1985, see also Aaquist \& Kwok 1990), showing several 
knots embedded in a faint elongated structure of $\simeq$ 3$\times$2 
arcsec$^2$ in size. The H$\beta$ surface brightness ($\log F({\rm H}\beta) 
= -11.34$, Acker et al. 1992) suggests that IC\,4846 has a high electron 
density. This is corroborated by the small 
[S~{\sc ii}]$\lambda$6717,$\lambda$6731 doublet ratio (Barker 1978; 
Acker et al. 1992) which reaches the limiting ratio for high electron 
density ($N_{\rm e} \gapprox 10^4$~cm$^{-3}$). Both its small angular size 
and high density indicate that IC\,4846 is a young PN. The line intensity 
ratios (Acker et al. 1992) point to a moderate excitation 
([N~{\sc ii}]$\lambda$6583/H$\alpha \simeq 0.07$, 
[O~{\sc iii}]$\lambda$4959/H$\beta \simeq 3.7$, and very faint 
He~{\sc ii} $\lambda$4686). {\it IUE} low-resolution UV spectra of the 
central star show a strong P-Cygni profile in the N~{\sc v} 
line\footnote{ Based on INES 
data from the {\it IUE} satellite.} with a terminal velocity of $\simeq$ 1300 
km~s$^{-1}$. The occurrence of a strong stellar wind in the central stars of 
PNe is indicative of their early evolutionary stage (Cerruti-Sola \& 
Perinotto 1985). 

The shaping processes of PNe occur early in their formation. Therefore, the 
signatures of these processes in the nebula are more recent and more likely 
to have withstood the fast nebular evolution in young PNe. This turns the 
observations of young PNe of chief importance in the understanding of their 
formation mechanisms. Accordingly, IC\,4846 was included in our 
observational program of compact young PNe (see Miranda 1999 and references 
therein), and narrow--band H$\alpha$ and [N~{\sc ii}] images and long--slit 
high--resolution spectra were obtained to study its structure and internal 
kinematics. The results reported in this paper evidence the existence of 
bipolar, precessing collimated outflows in IC\,4846. 

\section{Observations}

Direct images of IC\,4846 were obtained in 1997 July 24 with the Nordic 
Optical Telescope (NOT) at Roque de los Muchachos Observatory on La 
Palma\footnote{The Nordic Optical Telescope is operated on the Island
of La Palma by NOTSA in the Spanish Observatorio del Roque de los Muchachos
of the Instituto de Astrof\'{\i}sica de Canarias}. The detector was a 
LORAL 15 $\mu$m CCD at the HIRAC camera providing 
a plate scale of 0.11 arcsec\,pixel$^{-1}$. The filters were H$\alpha$ 
(${\lambda}_0~\simeq$ 6563 {\AA}, FWHM $\simeq$ 10 {\AA}), and [N~{\sc ii}] 
(${\lambda}_0~\simeq$ 6584 {\AA}, FWHM $\simeq$ 10 {\AA}). Exposure times 
were 900\,s in each case. The spatial resolution is $\simeq$ 1 arcsec. The 
images were bias subtracted and flat--fielded following standard procedures 
within MIDAS package.   

Long--slit echelle spectra of IC\,4846 were taken with the 
IACUB\footnote{The IACUB uncrossed echelle spectrograph was built in a 
collaboration between the IAC and the Queen's University of Belfast} 
spectrograph at the NOT in 1998 August 2. The detector was a Thompson 19 
$\mu$m CCD with 1024$\times$1024 pixels. A filter was used to isolate the 
H$\alpha$ and [N~{\sc ii}]$\lambda$6583 emission lines in the 9$^{th}$ 
echelle order. A Thorium--Argon spectrum was taken after each object 
spectrum for wavelength calibration purposes. Exposure time was 1800\,s for 
each spectrum. The observed position angles (PAs) were 11$^{\circ}$, 
30$^{\circ}$, 54$^{\circ}$, and 140$^{\circ}$ 
with the slit centered on the object. The dispersion was 2.5 {\AA}~mm$^{-1}$, 
and the achieved spectral resolution (FWHM) $\simeq$ 7.5 km\,s$^{-1}$. The 
spatial resolution is $\simeq$ 1.5 arcsec. Absolute errors in the radial 
velocity are $\pm$ 1.5 km\,s$^{-1}$, whereas relative errors (within each 
long-slit spectrum) are $\pm$ 0.5 km\,s$^{-1}$. The data were reduced 
following standard procedures for long--slit spectroscopy within the IRAF 
and MIDAS packages. Thereafter the [N~{\sc ii}]$\lambda$6583 emission line 
will be referred to as the [N~{\sc ii}] line.

\section{Results}

\subsection{General structures}

Figures~1 and 2 present grey--scale and contour maps, respectively, of 
IC\,4846 in the light of [N~{\sc ii}] and H$\alpha$. The morphology of 
IC\,4846, resolved in these images for the first time, displays a great 
wealth of structures. At the lowest intensity levels, the appearance of the 
nebula is elliptical with two bipolar features protruding at 
PA $\simeq$ 54$^{\circ}$. These features, denoted A1--A2 (Fig.~2), are curved 
filaments enhanced in [N~{\sc ii}]. A second, inner ellipse, observed in 
H$\alpha$ and [N~{\sc ii}] at intermediate intensity levels, can be 
recognized with its major axis at PA $\simeq$ 11$^{\circ}$. Two compact 
knots, B1--B2, are observed at the tips of this inner ellipse. The innermost 
regions of IC\,4846 are dominated in [N~{\sc ii}] at the highest intensity 
levels by two bright structures, C1--C2, which are not clearly distinguished 
in the H$\alpha$ image. An additional attached circular shell with a size of 
$\simeq$ 10$''$ can also be identified in the H$\alpha$ image. 

In order to emphasize the different structures of IC\,4846 described before, 
we have obtained deconvolved images using the Lucy-Richardson algorithm as 
implemented in the MIDAS package. Deconvolved H$\alpha$ and [N~{\sc ii}] 
images to a resolution of $\simeq$ 0.5 arcsec (obtained with 30 iterations) 
are also presented in Fig.~2. These images clearly show the structural 
richness of the nebula. In particular, the curvature and filamentary 
appearance of A1--A2 can easily be recognized as well as the compactness of 
B1--B2. C1--C2 appear as point--symmetric structures elongated perpendicular 
to the radial direction from the center. In H$\alpha$ the inner regions 
appear as an arc with an asymmetric intensity distribution. We note that the 
inner regions of the nebula resemble the morphology observed at 6\,cm 
(Kwok 1985). In addition, the inner shell presents a very low ellipticity. 

Figure~3 shows position--velocity (PV) contour maps of the high resolution 
long--slit spectra of the [N~{\sc ii}] emission line at PAs 
11$^{\circ}$, 54$^{\circ}$ and 140$^{\circ}$. The spectrum at PA 30$^{\circ}$ 
is similar to that at PA 11$^{\circ}$ and is not shown here. The bipolar 
features A1--A2, B1--B2 and C1--C2 can be identified in the PV maps with 
point-symmetric pairs of components. We also note that no {\it closed} 
velocity ellipse can be identified in the PV maps of the [N~{\sc ii}] 
emission line, which 
could be expected from the elliptical structures observed in the images. The 
H$\alpha$ spectra (not shown here) are dominated by emission from the inner 
regions. These regions show a tilt in the PV maps similar to the tilt of 
C1--C2 (Fig.3). A1--A2 are clearly distinguished because of their high 
radial velocity. B1--B2 seem to be present, too, although the H$\alpha$ line 
is not useful for a detailed kinematic study owing its large thermal width. 

A [N~{\sc ii}]/H$\alpha$ ratio map is also presented in Fig~1. A1--A2 and 
B1--B2 are highlighted due to their relatively strong [N~{\sc ii}] emission. 
C1--C2 can also be recognized, although the line ratio seems to have lower 
values. A quantitative estimate of [N~{\sc ii}]/H$\alpha$ should be worked 
out using the long-slit spectra in order to single out both spatially and 
kinematically the different components. PV maps of the [N~{\sc ii}]/H$\alpha$ 
ratio for PAs 11$^{\circ}$ and 54$^{\circ}$ are shown in Fig.~3. A mean 
value of [N~{\sc ii}]/H$\alpha$ in the nebula of $\simeq$ 0.06 is obtained, 
similar to that obtained by Acker et al. (1992). The three pairs present 
enhanced [N~{\sc ii}] emission, with mean [N~{\sc ii}]/H$\alpha$ values of 
$\simeq$ 0.8 (A1--A2), $\simeq$ 0.15 (B1--B2), and $\simeq$ 0.2 (C1--C2), 
which are a factor 2.5 to 13 higher than the mean ratio in the nebula. 
Furthermore, peak values 
of this ratio up to $\simeq$ 1.5 are found in A1--A2, and up to $\simeq$ 0.3 
in B1--B2 and C1--C2. Common to all three pairs is that 
[N~{\sc ii}]/H$\alpha$ peaks at larger distances from the center than the 
[N~{\sc ii}] emission peaks (see also Fig.~1).

Table~1 lists different parameters of the bipolar pairs as deduced from the 
[N~{\sc ii}] images and long--slit spectra. In order to deduce systemic 
velocities, we have first considered the centroid of each bipolar pair as 
representative of its own systemic velocity. The systemic velocity obtained 
for each pair is listed in Table~1. Secondly, we have obtained the 
centroid velocity of the 
[N~{\sc ii}] emission line as representative of the systemic velocity of 
the whole nebula. In order to avoid the contribution of the bipolar 
pairs, the centroid velocity was 
obtained from the middle point of the spatially integrated emission at 10\% 
level of the intensity peak. We obtain V$_{sys}$(centroid) $\simeq$ 168 
km\,s$^{-1}$ and note that an identical value is obtained if the middle point 
is considered at 5\% or 15\% levels of the intensity peak. The values for 
V$_{sys}$(centroid), V$_{sys}$(B1--B2) and V$_{sys}$(C1--C2) agree within 
the absolute errors with the value quoted by Bianchi (1992) and Durand et al. 
(1998) for the systemic velocity of IC\,4846. The systemic velocity of 
B1--B2 and C1--C2 are almost identical to each other; both are slightly 
blueshifted ($\simeq$ 3.5 km\,s$^{-1}$) with respect to the centroid 
velocity. In the case of A1--A2, its systemic velocity is noticeably 
blueshifted by $\simeq$ 10 km\,s$^{-1}$ with respect to the centroid velocity. 
We will discuss in \S\,4 possible interpretations for these systemic velocity 
shifts. In the following, radial velocities of each pair will be quoted 
with respect to its own systemic velocity.

\subsection{Properties of the collimated components}

IC\,4846 displays three pairs of structures, A1--A2, B1--B2, and C1--C2, 
which are highly evocative of collimated outflows. We describe as follows 
the properties of these components.

A1--A2 appear detached and off-axis ($\Delta$(PA) $\simeq$ 43$^{\circ}$) with 
respect to the major axis of the inner ellipse. Their orientation changes by 
$\simeq$ 5$^{\circ}$--10$^{\circ}$. A1--A2 can be traced along $\simeq$ 6 
arcsec, which represents a considerable fraction of the total nebular size. 
An upper limit of $\simeq$ 1 arcsec is obtained for their lateral size. The 
radial velocity systematically varies along A1--A2: a minimum velocity of 
$\simeq$ $\pm$ 36 km\,s$^{-1}$ is observed at the minimum angular distance; 
the velocity then steadily increases with the distance up to $\simeq$ $\pm$ 
52 km\,s$^{-1}$; finally, it slightly decreases at the tips of the features. 
The velocity width is small as compared with the velocity 
difference between A1 and A2 (see Table~1), which implies a high 
degree of collimation. According to their properties, A1--A2 are highly 
suggestive of a precessing bipolar jet.

B1--B2 present a rather moderate radial velocity and the velocity width is 
small and comparable to that of A1--A2 (Table~1). The small velocity 
width and the compactness indicate that B1--B2 are highly collimated 
outflows.

The radial velocity of C1--C2 does not depend on the PA (Fig.~3, Table~1), 
although some small variations ($\simeq$ 7 km\,s$^{-1}$) within C1--C2 are 
observed. Possible interpretations of this pair will be discussed in \S\,4.

\section{Discussion}

The observations of IC\,4846 have disclosed a complex structure composed of 
three different elements: (1) two apparent elongated shells with 
different orientations, (2) an attached circular shell and (3) three pairs of 
point-symmetric components at different orientations with typical properties 
of collimated outflows. In the following we discuss about the nature, 
models and possible interpretations of these components.

As already mentioned, there is no kinematic evidence for the elliptical 
shells in the PV maps of the [N~{\sc ii}] line, although the extended 
emission observed in the PV maps may be related to these shells. We also 
note that the major axis of the outer ellipse is only defined by A1--A2. 
In fact, without A1--A2 the outer ellipse could not be recognized. Images 
at higher resolution would be necessary to elucidate the precise geometry 
of these shells. While A1--A2 and B1--B2 can be interpreted as precessing 
bipolar outflows, the interpretation of the pair C1--C2 is somewhat 
difficult. Its kinematic properties and the elongation perpendicular to the 
radial direction suggest that C1--C2 could represent collimated ejections in 
which the orientation of the ejection axis has remarkably changed over the 
ejection time. However, we cannot completely rule out that C1--C2 represents 
a ring-like structure containing two bright opposite regions. The attached 
circular shell probably represents 
a previous slow wind ejected by the AGB progenitor of IC\,4846. It is similar 
to other attached shells observed in PNe and its formation may be related 
to the dynamics of ionization fronts (see Mellema 1995 and references therein).

In the scenario that A1--A2 are precessing bipolar jets, we can obtain 
estimates for the 
inclination angle and velocity by assuming that the expansion velocity is 
constant and that the variation of the radial velocity is due to a 
systematic change of the angle between the observer and the precessing jet 
along a precession cone with aperture angle 
$\simeq$ 5$^{\circ}$--10$^{\circ}$. The 
observed minimum and maximum radial velocities are assumed to be related to 
the far and rear side of the precession cone, respectively. Using the model 
for a biconical flow described by Miranda \& Solf (1992), we obtain an 
inclination angle for the precession axis of 
$\simeq$ 13$^{\circ}$--26$^{\circ}$ with respect to the plane of the sky, 
and a velocity of $\simeq$ 100--190 km\,s$^{-1}$. The (mean) kinematic age 
is $\simeq$ 200$\times$D(kpc) yr, where D is the distance to IC\,4846, which 
is extremely uncertain with published 
values between 2 and 11 kpc (see Acker et al. 1992). The images also show that 
A1--A2 protrude from the circular shell. This result, the increase of the 
[N~{\sc ii}]/H$\alpha$ ratio at larger distances and the decrease of the 
radial velocity at the tips of A1--A2 suggest an interaction between the 
jet fronts and the circular shell, in a similar way to that reported 
in, e.g., NGC\,6891 (Guerrero et al. 2000). 

It is worthwhile to emphasize the extraordinary similarity between 
IC\,4846 and the well-known PN NGC\,6543 (see Miranda \& Solf 1992 and Reed 
et al. 1999). The following structural components are common to these two 
PNe (we will use for NGC\,6543 the nomenclature of Miranda \& Solf 1992): 
\begin{enumerate}
\item {\em Basic elongated shell(s).} In NGC\,6543 two shells are observed, an 
outer bipolar shell with a bright ring between the lobes and an inner 
elliptical one. In IC\,4846 two elongated shells are suggested by the images 
(but see above). The possible existence of an equatorial ring should be 
confirmed by new data.
\item {\em Bipolar knots at the tips of the shell.} The knots FF' in 
NGC\,6543 correspond to B1--B2 in IC\,4846.  
\item {\em Precessing bipolar jets.} The bipolar jets JJ' in 
NGC\,6543 correspond to A1--A2 in IC\,4846. The similarity between JJ' and 
A1--A2 is extraordinary. Both appear off-axis with respect to the orientation 
of the basic shell and are curved, with a change of orientation of 
5$^{\circ}$--10$^{\circ}$. The radial velocity presents an identical variation 
along the jets with a steady increase up to a maximum radial velocity and a 
small decrease at their tips. Finally, an expansion velocity of 
$\simeq$ 100--200 km\,s$^{-1}$ is obtained in both cases. 
\item {\em Point-symmetric components perpendicular to the radial 
direction.} C1--C2 in IC\,4846 could correspond to DD' in NGC\,6543.
\end{enumerate}

There are some differences between both objects, but these are most likely 
attributable to different orientations, ages and distances. Whatever these 
differences, the conspicuous kinematic and structural similarities lead us 
to conclude that NGC\,6543 and IC\,4846 have shared a similar formation 
history.  

An interesting result in IC\,4846 is the systemic velocity shifts among  
the different components and the centroid of the nebula. In the 
case of B1--B2 and C1--C2, the shifts are only slightly larger than the 
relative errors and we cannot conclude whether they are real. In the case 
of A1--A2, however, the shift is noticeable and we consider it real. Both 
the remarkable spatial and velocity symmetry observed in A1--A2 exclude that 
this shift may be ascribed to different ejection velocity, changes 
(deceleration) in the velocity in one of the components of the pair, 
different ejection angle or a combination of the three. A more plausible 
explanation is that the systemic velocity of A1--A2 is indeed intrinsically 
different from that of the other components and, in particular, from that 
of the emission centroid. Similar shifts in the systemic velocity of 
collimated outflows and shell have been recently reported in the PN 
Hu\,2-1 (Miranda et al. 2000). Furthermore, both PNe share the presence of 
two systems of collimated outflows at different orientations, one of them 
projected along the nebular major axis and one off-axis, suggesting a 
parallel formation history. Moreover, inspection of the kinematic data 
of NGC\,6543 (see Miranda \& Solf 1992) suggests that a difference of 
$\sim$ 7 km\,s$^{-1}$ could exist between the systemic velocity of JJ' 
and the systemic velocity of the nebula as defined by the rest of components. 
In the case of Hu\,2-1, the shifts of systemic velocity have been 
interpreted as a direct evidence of orbital motion of the ejection source 
in a binary central star (Miranda et al. 2000). The velocity shifts detected 
in IC\,4846 fit also into this framework. If A1--A2 are ejected from a star 
in a binary system, the space velocity of the A1--A2 will contain an 
additional 
velocity component due to the orbital motion, besides to the component due 
to the own ejection velocity. Accordingly, the systemic velocity of A1--A2 
will be shifted from that of the emission centroid of the nebula (see 
Miranda et al. 2000 for details). We note that the same would apply to 
B1--B2 and C1--C2 if their observed velocity shifts were real.  

The systemic velocity shift of A1--A2 (10 km\,s$^{-1}$) is very similar to 
that found in Hu\,2-1 (8 km\,s$^{-1}$) and in NGC\,6543 (7 km\,s$^{-1}$). 
Consequently, the orbital parameters of their binary central stars should be 
rather similar as well, with an orbital separation $\leq$ 30 AU and an 
orbital period $\leq$ 100 yr, thus linking the central stars of these 
nebulae to symbiotic stars. These results strongly suggest that the central 
star of IC\,4846, similarly to that of Hu\,2-1 and NGC\,6543, is a binary 
star in which mass transfer from a red giant leads to the formation of an 
accretion disk about a secondary (Mastrodemos \& Morris 1998, 1999), from 
which collimated ejections may be generated.

\section{Conclusions}

Narrow--band imaging and high--resolution spectroscopy of the PN IC\,4846 
in the [N~{\sc ii}]$\lambda$6583 and H$\alpha$ lines have revealed a complex 
structure similar in many respects to that of NGC\,6543. The most remarkable 
components in IC\,4846 are three pairs of point-symmetric structures oriented 
at different directions. Two (and probably the three) pairs represent 
collimated bipolar outflows. The kinematics and morphology of the pair 
observed at PA~54$^{\circ}$ (A1--A2) are highly suggestive of a precessing 
jet moving at $\simeq$ 100--200 km\,s$^{-1}$. Furthermore, the 
velocity shift between their systemic velocity and that of the nebula can be 
interpreted as a result of orbital motion of the ejection 
source in a binary central star. This kind of systemic velocity differences 
has also been observed in Hu\,2-1 and could be present in NGC\,6543. Thus,  
the formation of IC\,4846 (and Hu\,2-1 and NGC\,6543) could be related to 
physical processes in the own evolution of interacting binary stars.

\vspace{1.5cm}

\noindent {\bf Acknowledgments}   

We thank our referee for his/her valuable comments which have improved the 
presentation of the paper. LFM and JMT are supported partially by 
DGESIC PB98-0670-C02 and Junta de Andaluc\'{\i}a (Spain). MAG is 
supported partially by DGESIC of the Spanish Ministerio of Educaci\'on 
y Cultura.
                                                                   
\newpage

\noindent {\bf References}  

\noindent Aaquist O.B., Kwok S., 1990, A\&AS, 84, 229

\noindent Acker A., Ochsenbein F., Stenholm B., Tylenda R., Marcout J., 
Schohn C., 1992, Strasburg-ESO Catalogue of Galactic Planetary Nebulae. ESO,
Garching           

\noindent Barker T., 1978, ApJ, 219, 914

\noindent Bianchi L., 1992, A\&A, 260, 314 

\noindent Cerruti--Sola M., Perinotto M., 1985, ApJ, 291, 237

\noindent Durand S., Acker A., Zijlstra A., 1998, A\&AS, 132, 13  

\noindent Guerrero M.A., Miranda L.F., Manchado A., V\'azquez R., 2000, 
MNRAS, 313, 1

\noindent Kwok S., 1985, AJ, 90, 49

\noindent Mastrodemos N., Morris M., 1998, ApJ, 497, 303  

\noindent Mastrodemos N., Morris M., 1999, ApJ, 523, 357 

\noindent Mellema G., 1995, MNRAS, 277, 173

\noindent Miranda L.F., 1999, in ASP Conf. Ser., vol.188, p. 257 

\noindent Miranda L.F., Solf J., 1992, A\&A, 260, 397

\noindent Miranda L.F., Torrelles J.M., Guerrero M.A., V\'azquez R., 
G\'omez Y., 2000, MNRAS, in press

\noindent Reed D.S., Balick B., Hajian, A.R., Klayton T.L., Giovanardi, S., 
Casertano S., Panagia N., Terzian Y., 1999, AJ, 118, 2430

\newpage

\begin{center}

{\bf Table\,1: Parameters of the [N~{\sc ii}] bipolar pairs in IC\,4846$^a$}

\begin{tabular}{lcccccc}
\noalign{\smallskip}
\hline
\noalign{\smallskip}
Pair & $\delta$V$_r$$^b$ & $\Delta$V$^c$ & V$_{\rm sys}$(LSR)$^d$ & 
$\delta$V$_{\rm sys}$$^e$ & $\delta$X$^f$ & PA$^g$ \\  
\noalign{\smallskip}
& (km\,s$^{-1}$)  & (km\,s$^{-1}$) & (km\,s$^{-1}$) & (km\,s$^{-1}$) 
& (arcsec) & (degrees)\\
\hline
\noalign{\smallskip}  
A1--A2   & 96 & 20 & 158 & $-$10 & 7.4 & 54 \\
B1--B2   & 26 & 18 & 164 & $-$4 & 5.4 & 11 \\
C1--C2   & 38 & 22 & 165 & $-$3 & 1.8 & 25 \\
\noalign{\smallskip}
\hline
\end{tabular}

\end{center}

\vspace{0.3cm}

\noindent$^a$ Obtained from the images and long--slit [N~{\sc ii}] spectra 
shown in Figs.~1 and 3, respectively. \\ 
$^b$ Separation in radial velocity between intensity peaks of the components 
of each pair. \\
$^c$ Velocity dispersion (FWHM) corrected of instrumental resolution. \\ 
$^d$ Systemic velocity. \\
$^e$ Maximum difference between V$_{\rm sys}$(LSR) for each pair and 
V$_{\rm sys}$(LSR) for the centroid of the [N~{\sc ii}] emission line = 
168 km\,s$^{-1}$ \\
$^f$ Projected angular separation between the intensity peak of the components 
of each pair . \\
$^g$ Orientation (position angle).

\newpage

\begin{figure}
\plotone{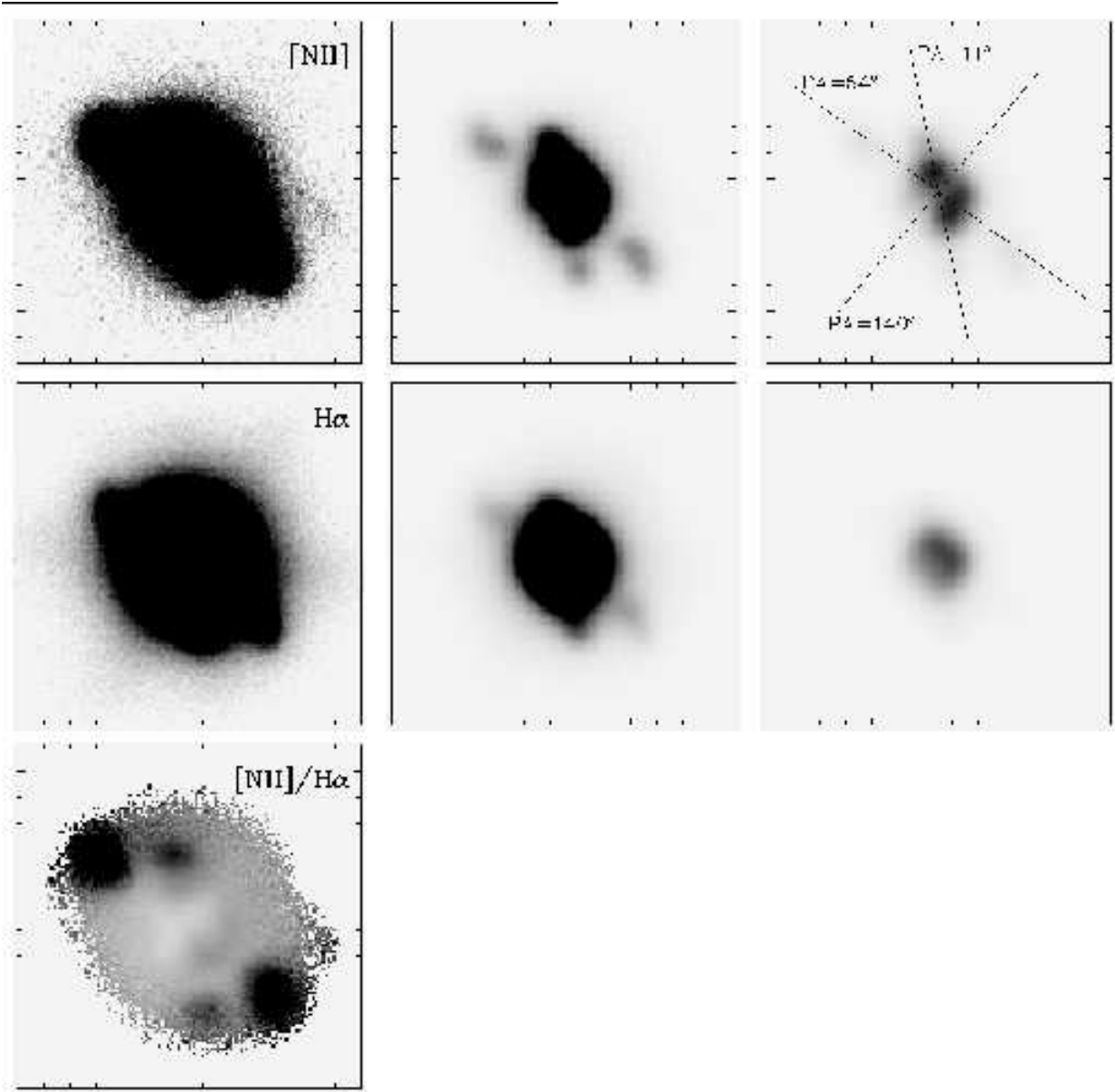}
\caption{Grey--scale maps of the [N~{\sc ii}]$\lambda$6583 {\it (upper row)}, 
H$\alpha$ {\it (middle row)} and [N~{\sc ii}]$\lambda$6583/H$\alpha$ 
{\it (lower row)} images of IC\,4846. The [N~{\sc ii}] and H$\alpha$ images 
are presented in three different 
contrasts in order to show the faint and bright nebular regions. The 
observed slit positions at PAs 11$^{\circ}$, 54$^{\circ}$ and 140$^{\circ}$ 
are indicated in one of the [N~{\sc ii}] maps. In the 
[N~{\sc ii}]/H$\alpha$ map, black indicates stronger [N~{\sc ii}] 
emission. North is up, east to the left. The size of the field shown is 
14$\times$14 arcsec$^2$.}
\end{figure}

\newpage

\begin{figure}
\plotone{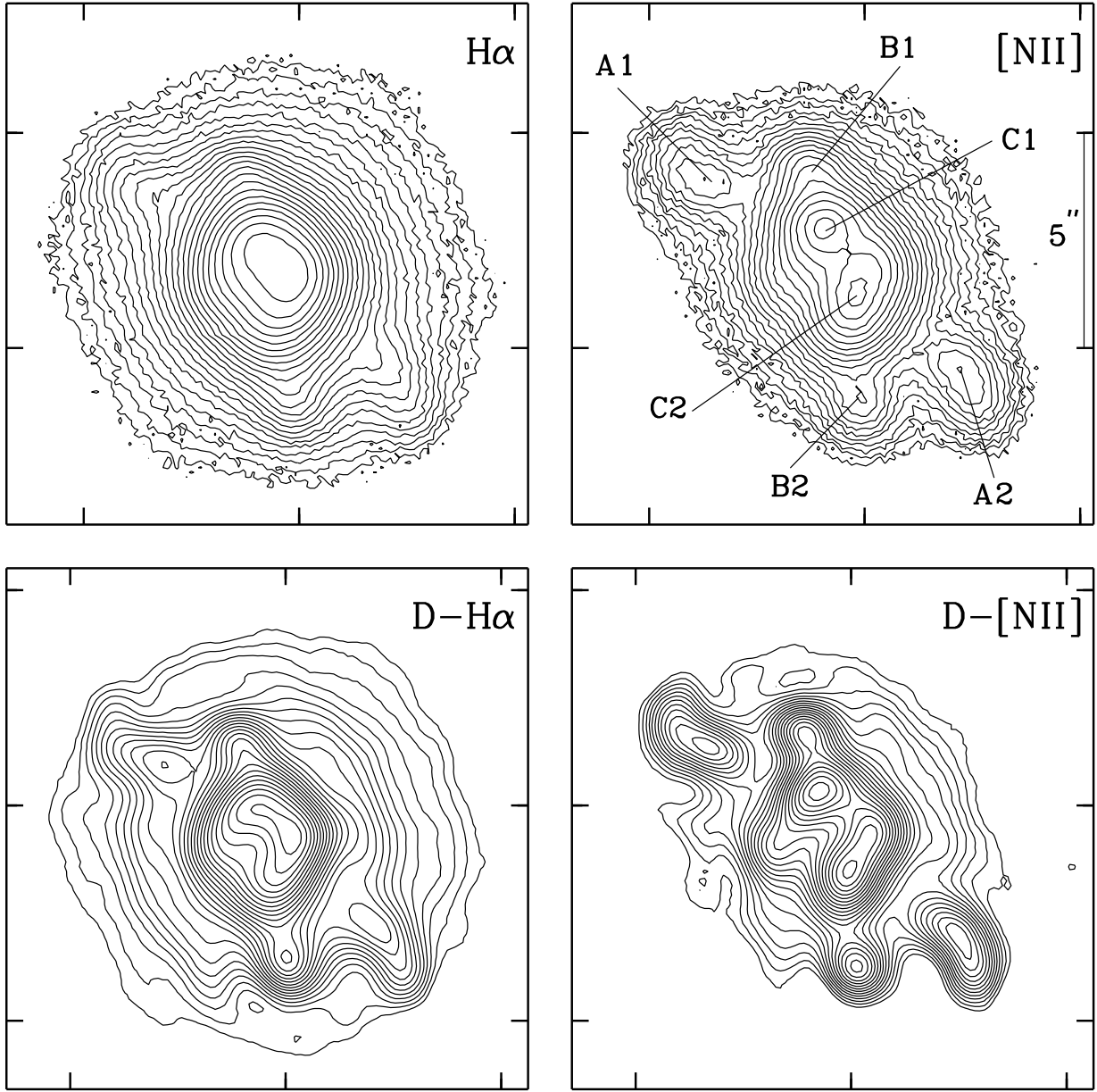}
\caption{{\it (upper row)} Contour plots of the images of 
IC\,4846 presented in Fig.~1. The pairs of bright [N~{\sc ii}] components are 
labeled A1--A2, B1--B2 and C1--C2. {\it (lower row)} Deconvolved H$\alpha$ and 
[N~{\sc ii}] images at a resolution of 0.5 arcsec. The contours are 
logarithmic separated by a factor 1.32 in intensity.}
\end{figure}   

\newpage

\begin{figure}
\plotone{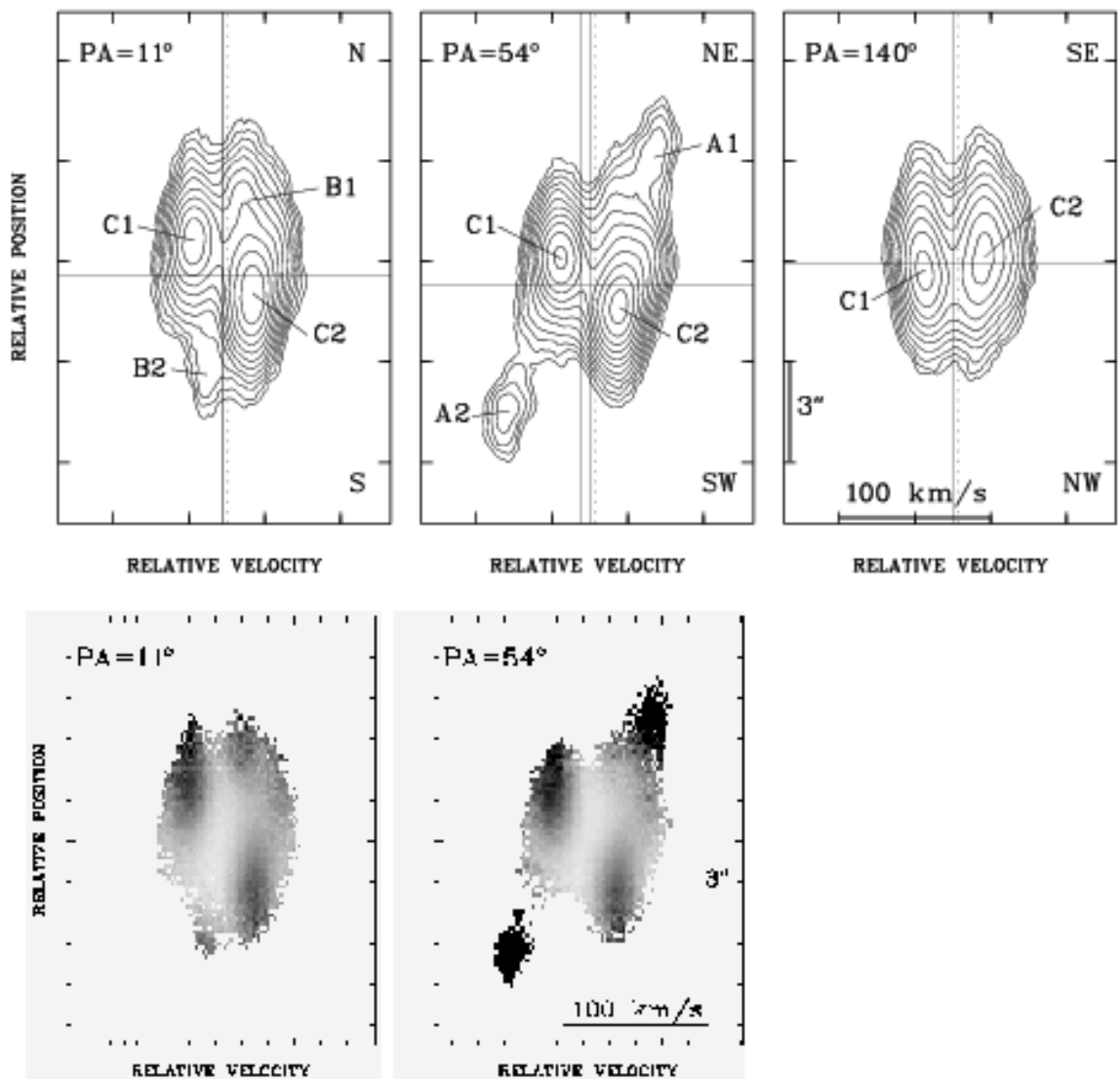}
\caption{{\it (upper row)} Contour plots of the long--slit 
[N~{\sc ii}]$\lambda$6583 spectra at three different PAs (upper left) in a 
position--velocity representation. The contours are logarithmic separated by 
a factor 1.41 in intensity. The identified components are labeled (see 
Fig.~2). The horizontal line corresponds to the mean position of C1--C2. The 
vertical lines correspond to the systemic velocity of the different pairs 
(continuous lines) and to that of the emission centroid (dashed line) (see 
text). {\it (lower row)} Grey--scale maps of the 
[N~{\sc ii}]$\lambda$6583/H$\alpha$ ratio deduced from the long--slit 
spectra of IC\,4846 at PAs 11$^{\circ}$ and 54$^{\circ}$. Black indicates 
stronger [N~{\sc ii}] emission.}
\end{figure}

\end{document}